\documentclass[reprint,superscriptaddress,showpacs,amssymb,floatfix]{revtex4-1}
\usepackage[final]{graphicx}
\usepackage{amsmath}
\usepackage{siunitx}

\begin{document} 

\title{The optical M\"{o}bius strip cavity: Tailoring geometric phases and far fields}

\author{Jakob Kreismann}
\email{jakob.kreismann@tu-ilmenau.de}
\affiliation{Institute for Physics, Theoretical Physics II/Computational Physics Group, Technische Universit\"{a}t Ilmenau, Weimarer Stra\ss{}e 25, 98693 Ilmenau, Germany}
\author{Martina Hentschel}
\affiliation{Institute for Physics, Theoretical Physics II/Computational Physics Group, Technische Universit\"{a}t Ilmenau, Weimarer Stra\ss{}e 25, 98693 Ilmenau, Germany}

\date{\today}

\begin{abstract}
\textbf{The M\"{o}bius strip, a long sheet of paper whose ends are glued
	together after a $180^{\circ}$ twist, has remarkable geometric and topological
	properties. Here, we consider dielectric M\"{o}bius strips of finite width
	and investigate the interplay between geometric properties and resonant light propagation. 
	We show how the polarization dynamics of the electromagnetic wave depends on the topological properties,
	and demonstrate how the geometric phase can be manipulated between $0$ and $\pi$ through the system geometry. The loss of the M\"{o}bius character in thick cavities and
	for small twist segment lengths allows one to manipulate the polarization dynamics and the far-field emission, and opens the venue for applications.}
\end{abstract}

\pacs{
42.55.Sa, 
42.60.Da, 
42.25.Ja  
}

\maketitle

\begin{figure}
	\centering
	\includegraphics[width=8.5cm]{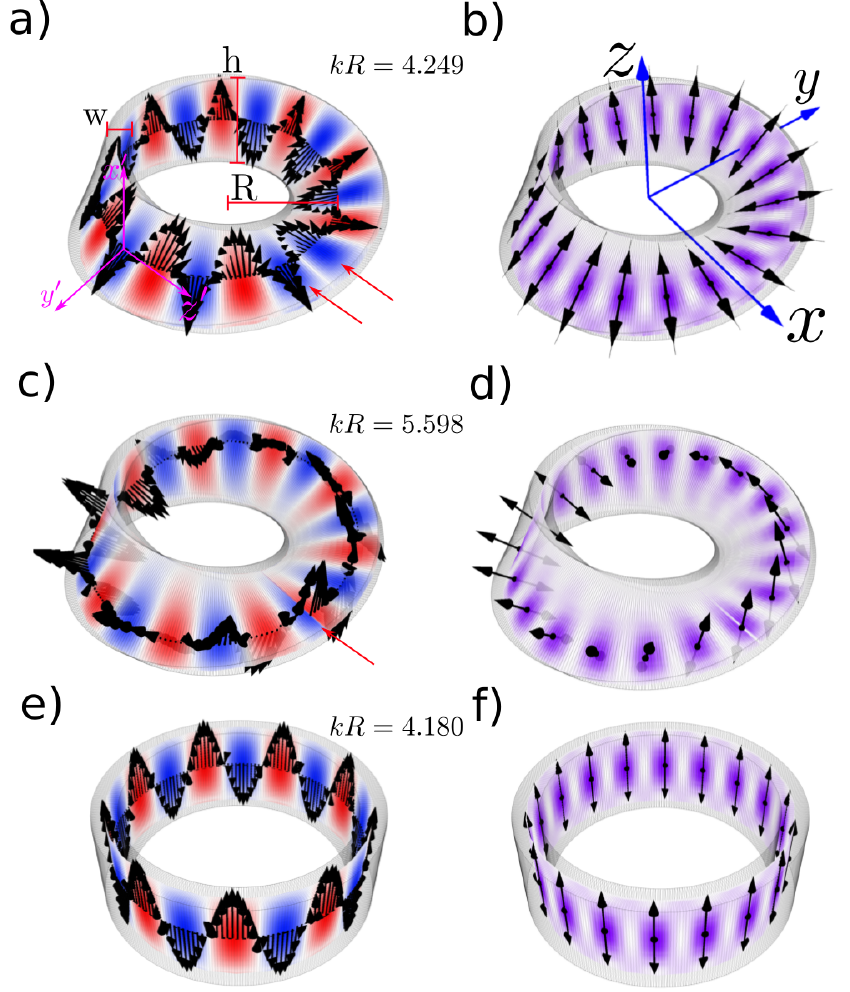}
	\caption{Electric field (left column) and intensity plots (right column) of the M\"{o}bius strip cavity (a - d) and of an usual ring cavity (e and f). $kR$ is the dimensionless resonance frequency. Left column: Black arrows represent the electric field vector of a generalized WG-type mode for TE and TM polarization, and TE polarization in the ring cavity taken at a certain time. Blue and red areas illustrate the projection of the electric field vector parallel (a and e) and perpendicular to (c) the tube wall; blue (red) represents a negative (positive) value of this projection. Two adjacent blue areas (a) result from the non-orientabilty of the strip. Right column: Intensity plots where black double-headed arrows  represent the polarization orientation calculated over one oscillation period in time.
	}
	\label{fig:f1}
\end{figure}

Optical microcavities have attracted a lot of interest in the past two decades ~\cite{Vahala2003,Chang1996}. Especially, three-dimensional (3D) optical microcavities are an interesting object of research, see e.g.~\cite{Kreismann2017}. Unlike the two-dimensional case, the polarization is not fixed to transvere magnetic (TM) or transverse electric (TE). Rather, it may change, and the resulting polarization evolution is a characteristic system property. 
The polarization dynamics of light in terms of spin-orbit interaction has been investigated in recent years, see e.g.~\cite{Bliokh2015,Cardano2015,Slussarenko2016,Ma2016} where the polarization evolution was determined by geometric phases arising from optical spin-orbit interaction. These spin-orbit interactions are based on light propagating in smoothly inhomogeneous or anisotropic materials. Earlier works demonstrate the appearance of geometric phases in coiled fibers~\cite{Tomita1986,Chiao1986,Haldane1986}, and the Pancharatnam phase of beams whose polarization state undergoes certain changes~\cite{Pancharatnam1956}, verified in Mach-Zender interferometry~\cite{Jiao1989,Chiao1988}.\\
In this paper, we show the presence and the consequences of the Pancharatnam phase in a 3D microcavity, namely the dielectric M\"{o}bius strip. 
The M\"{o}bius-strip is a fascinating and special surface, not only from the mathematical point of view. It can be created by joining the ends of a half-twisted strip to form a loop. The M\"{o}bius-strip is a surface with only one side and non-orientable, hence it is not possible to make a consistent choice of a surface  normal vector at every point.\\
M\"{o}bius strips have been realized in a wire-loaded copper cavity~\cite{Mobius-Copper}, in a electric circuit resonator~\cite{Mobius-elecCircuit}, by coupled resonators~\cite{Mobius-analog}, in twisted ribbon crystal materials~\cite{Mobius-crystal}, as plasmonic resonances in nano metallic strips~\cite{Mobius-plasmon}, or in metamaterials~\cite{Mobius-metaMat}. Quantum mechanics on a M\"{o}bius ring has been studied in~\cite{MobiusQM}. Furthermore, optical (cavity-free) polarization M\"{o}bius-strips  were investigated using focused light~\cite{OptMob2010,OptMob2011,OptMob2014,OptMob2015} or light scattered by nanoparticles~\cite{OptMob2017}.  A theoretical investigation of plasmonic resonances in metallic M\"{o}bius nanorings and dielectric M\"{o}bius nanorings has been reported in~\cite{Yin2017} and~\cite{Li2013}.\\ 
Here, we will investigate a pure dielectric M\"{o}bius-strip cavity, discuss how the geometry has to be modified in order to tune the topological phase, and discuss the far-field properties of these modified structures. We consider a cavity of radius $R=\SI{2.29}{\micro\meter}$, height $h=\SI{1.75}{\micro\meter}$, and wall thickness $w=\SI{0.4}{\micro\meter}$ with a homogeneous and isotropic refractive index $n=3.3$ (c.f. FIG.~\ref{fig:f1}). We choose the parameters such that a thin strip with $w < h < R$ results. Furthermore, we are interested in resonances where the medium wavelength is smaller than the wall thickness, $w<\lambda/n$. We perform full 3D finite difference time domain (FDTD) wave calculations in order to compute the electricmagnetic field, its polarization direction, the light intensity and the far fields using the free software package MEEP~\cite{MEEP}.

\textbf{Polarization dynamics in the standard M\"{o}bius strip}.
A three-dimensional pure dielectric ring cavity can support TE-like and TM-like polarization modes~\cite{Jackson}. In the following analysis, we focus on the TE-like modes
where the electric field $\textbf{E}$ is almost transverse to the propagation direction $\textbf{k}$ ($\textbf{k}\cdot \textbf{E}=0 $) and remains parallel to the wall of the M\"{o}bius ring cavity during its evolution along the strip. An example of the resulting, generalized whispering-gallery-(WG-)type resonance is shown in Fig.~\ref{fig:f1} a). In a local coordinate system $(x',y',z')$ as depicted in Fig.~\ref{fig:f1} a), where $(x',y')$ represent the transverse coordinates (in radial, $R$, and vertical direction) and $z'$ is the longitudinal coordinate (along the propagation direction),  
the polarization orientation of the electric field $\textbf{E}$ can be described by the complex unit vector 
\begin{equation}
\textbf{p}=(p_{\text{x}'},p_{\text{y}'}),
\end{equation}
\noindent where $\textbf{p}^\ast\cdot\textbf{p}=1$. The longitudinal component $p_{\text{z}'}$ is omitted due to the almost transverse electric field vector. In case of linearly polarized light, the components $p_{\text{x}'}$ and $p_{\text{y}'}$ are purely real.\\
The following analysis is motivated by Pancharatnam\textquoteright s work~\cite{Pancharatnam1956}. 
Later, Berry~\cite{Berry1987} expressed Pancharatnam\textquoteright s contribution in a quantum-mechanical language which we will adopt here. As pointed out by Berry~\cite{Berry1987},  this representation of a polarized wave is equivalent to a vector $|\Psi\rangle$ describing a two-component spinor
\begin{equation}
|\Psi\rangle = \left(\Psi_+,\Psi_- \right)= \frac{1}{\sqrt{2}} \left(p_{\text{x}'}+i p_{\text{y}'},p_{\text{x}'}-i p_{\text{y}'} \right).
\end{equation}

\begin{figure}
	\centering
	\includegraphics[width=8.5cm]{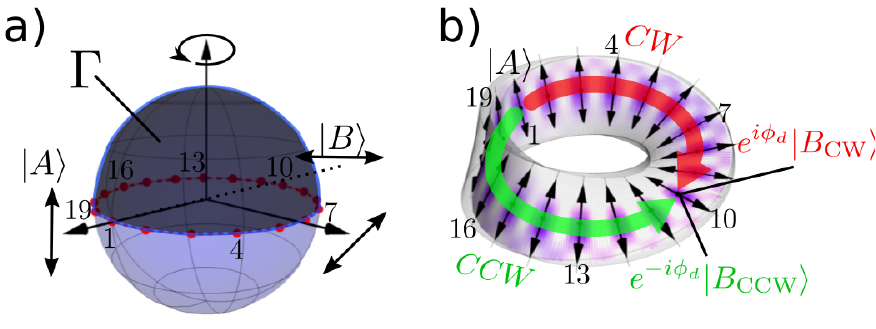}
	\caption{ \textbf{a)} Poincar\'{e} sphere of polarization. States $|A\rangle$ and $|B\rangle$ represent vertical and horizontal linear polarization. Unlabeled state between $|A\rangle$ and $|B\rangle$ on the equator is $45^{\circ}$ inclined horizontal linear polarization. The poles represent circular polarization. The red dots on the equator correspond to the intensity maxima (every third is numbered, c.f. FIG~\ref{fig:f2} b).~$\Gamma$ is the solid angle spanned by this evolution. \textbf{b)} Initial state $|A\rangle$ propagates in CW and CCW direction, and interferes at the opposite side.  }
	\label{fig:f2}
\end{figure}

\noindent Each such $|\Psi\rangle$ can be illustrated as a point on the so-called Poincar\'{e} sphere with spherical coordinates $(\theta,\phi)$ that describe the phase difference between $p_{x'}$ and $p_{y'}$, $\theta=\text{arg}(p_{x'}/p_{y'})+\pi/2$, and the orientation, $\phi$, of the principal axis of the polarization ellipse with respect to $x'$-axis, respectively. 
The poles of this sphere, $\theta=0 (\pi)$ correspond to vanishing  
$\Psi_+ (\Psi_-) $ components, and represent left (right) circularly polarized light. Points on the equator, $\theta=\pi/2$, indicate linear polarization into various directions $\phi$ where always 
$|\Psi_+|=|\Psi_-|$.\\
As the light propagates in the M\"{o}bius-strip, the polarization orientation changes from vertical at $|A\rangle$ in FIG.~\ref{fig:f2} to horizontal $|B\rangle$ and back again. This evolution is represented by a closed curve along the equator of the Poincar\'{e} sphere, as depicted in Fig.~\ref{fig:f2} a). Most importantly, we notice the number of intensity maxima $N=19$ to be an odd number which corresponds to an fractional azimuthal mode number $m=N/2=9.5$. To understand this unusual behavior, we recall the condition of resonance formation in a ring-like resonator. It is well known that the formation of a resonance requires a phase difference of $2\pi$ of two interfering waves, e.g., the clockwise (CW) and counter-clockwise (CCW) propagating waves. 
Let us consider the horizontally polarized initial state $|A\rangle$, cf.~Fig.~\ref{fig:f2} b). Starting from this state, waves propagate in CW and CCW direction. After half a round trip along the cavity (indicated by the green and red curved arrow, cf.~Fig.~\ref{fig:f2}b), both waves interfere and the resulting intensity $I$ is given by: 
\begin{align}
\begin{split}
I & = \left| e^{-i\phi_d}|B_{\text{CCW}}\rangle + e^{i\phi_d}|B_{\text{CW}}\rangle \right|^2 \\
& = 2 + 2 |\left\langle B_{\text{CCW}}|B_{\text{CW}}\right\rangle| \cos\left(\text{arg}( \left\langle B_{\text{CCW}}|B_{\text{CW}}\right\rangle) - 2\phi_d \right),\\
\end{split}
\end{align}
\begin{figure}
	\centering
	\includegraphics[width=8.5cm]{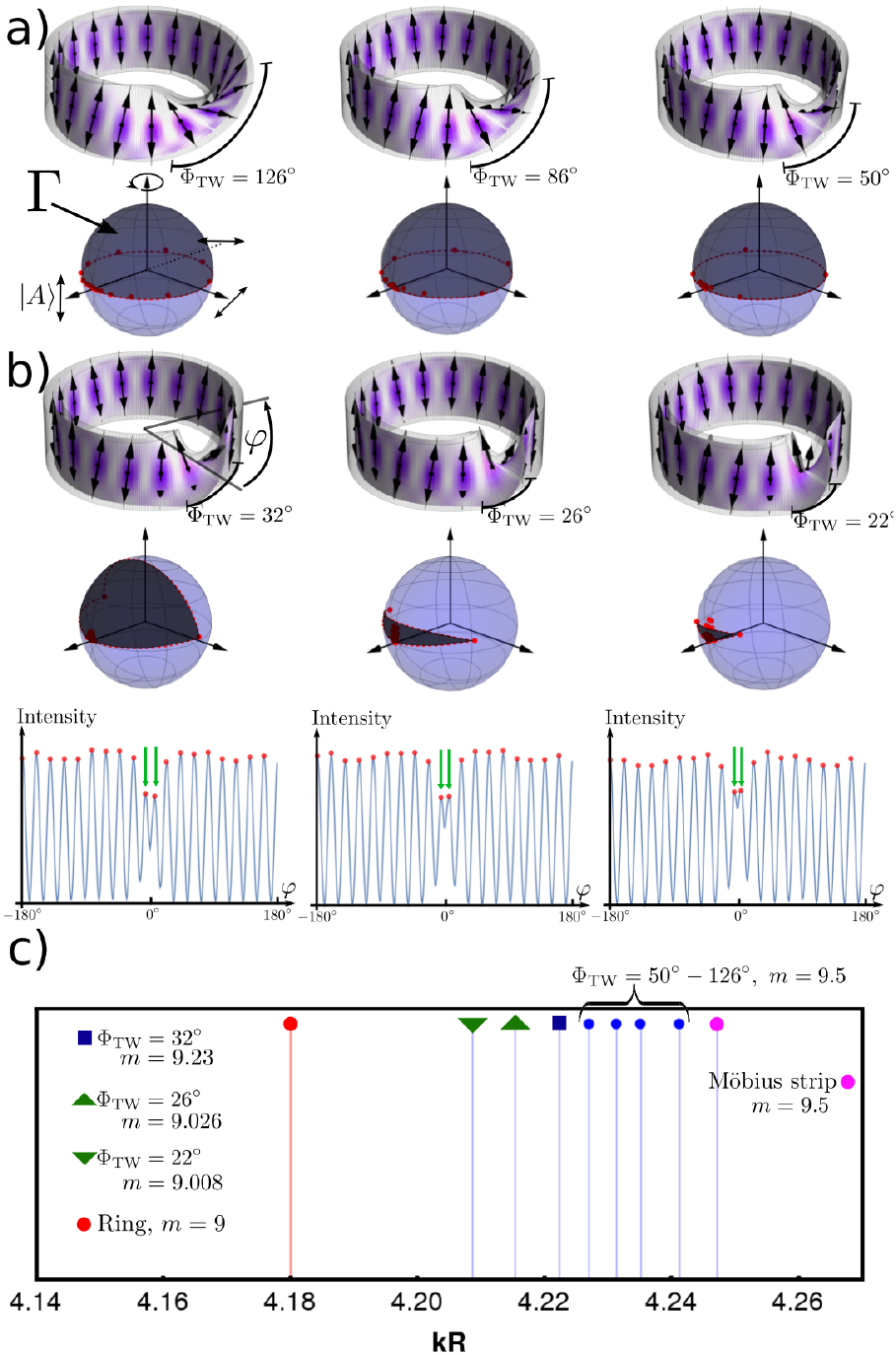}
	\caption{ \textbf{a)} Upper row shows three M\"{o}bius strips with different lengths of the twisted segments as indicated, and the intensity plots of the generalized WG-type modes with the corresponding polarization orientation. For all three cases, the twisting region is big enough for the solid angle to span the full hemisphere (lower insets). \textbf{b)} When the lengths of twisted segments are shorter, the polarization change will happen in a contracted region and therefore faster. The spanned solid angle is much less than $2\pi$. The lower row insets display the corresponding maximum intensity profiles.  The two maxmia around $0^{\circ}$ begin to merge as the solid angle decreases. \textbf{c)} Resonance wave number $kR$ and corresponding azimuthal mode number $m=q+\Gamma/(4\pi)$ for M\"{o}bius strips with varying geometry (blue) and a corresponding ring strip resonator (red).
		}
		\label{fig:f3}
\end{figure}

\noindent where $\phi_d=n k_{\text{r}} R \pi$ is the dynamical phase of half a round trip. It is related to the azimuthal  number $m$ by $ m = nk_{\text{r}}R $ that counts the number of medium wavelengths $\lambda/n$ to match the length of resonator $2\pi R$. In order to realize the resonance condition (maximum intensity), the argument of the cosine  has to be an integer multiple $q$ of $2\pi$: 
\begin{equation}
\text{arg}( \left\langle B_{\text{CCW}}|B_{\text{CW}}\right\rangle) - 2\phi_d=q2\pi.
\end{equation}
\textbf{Geometric phase}. As pointed out by Pancharatnam~\cite{Pancharatnam1956} and reformulated by Berry~\cite{Berry1987}, $\text{arg}( \left\langle B_{\text{CCW}}|B_{\text{CW}}\right\rangle)=\Gamma/2$ is the geometric phase equal to the half of the solid angle spanned by the curve on the Poincar\'{e} sphere. Inserting the dynamical phase $\phi_d$ and rearranging yields the azimuthal mode number $m$:
\begin{equation}
m=q+\frac{\Gamma}{4\pi}.
\label{equ:m}
\end{equation}

\noindent The azimuthal mode number is a function of the solid angle $\Gamma$ on the Poincar\'{e} sphere and can take integer, half-integer or arbitrary-fractional values if $\Gamma$ equals $0$ (usual ring resonator), $2\pi$ (usual M\"{o}bius strip, cf.~Fig.~\ref{fig:f2}) or any other angle (modified M\"{o}bius strip, cf.~Fig.~\ref{fig:f3}), respectively.\\ 
\textbf{M\"{o}bius ring cavities with half-integer $m$}. For the moment, we discuss the case of $\Gamma=2\pi$ where the upper hemisphere of the Poincar\'{e} sphere is spanned by the polarization evolution, cf.~Fig.~\ref{fig:f2}. Hence, the azimuthal mode number $m$ is half-integer $m=q+\frac{1}{2}$ and as a consequence, a standing wave must fit an half-integer number of medium wavelengths along the circumference of the M\"{o}bius strip: $2\pi R=(q+1/2)\lambda/n$. The number of intensity maxima is given by $N=2m=2q+1$ which yields an odd number and readily explains our above findings, c.f. FIG.~\ref{fig:f1}. In other words, the change of the polarization orientation during the propagation represents an adiabatic parallel transport of linear polarization which causes a geometric phase. This geometric phase is quantified by (one-half of) the corresponding solid angle in the parameter space, in our case the wave polarization parameter space which is governed by the same algebra as spin-$1/2$ particles~\cite{Berry1987}.\\ 
More importantly, the additional geometric phase of $\pi$ does not cause a phase jump as it might seem at the first sight in FIG.~\ref{fig:f1} a) and c). Rather, the two adjacent blue areas and the blue-red area result from the non-orientabilty of the strip. The TE-like (TM-like) mode in a) (c)) has a minimum (maximum) at the position where the strip is glued together and therefore we see two areas with the same color (one blue-red area). Note that, the electric field vectors (black arrows) show indeed a continuous evolution without any phase jump. \\
We would like to point out once more that all these optical modes confined in a thin and unperturbed M\"{o}bius strip have half-integer azimuthal mode number $m$. In contrast, quantum mechanical modes on a M\"{o}bius strip (as a solution of the Schr\"{o}dinger equation, see~\cite{MobiusQM}) exhibit both integer and half-integer azimuthal quantum numbers. In our case, the optical states undergo parallel transport of polarization which is different from the quantum mechanical states where the azimuthal quantum number depends on an even or odd transverse quantum number~\cite{MobiusQM}. Furthermore, studies of propagation of polarized light in twisted waveguides show that the geometric phase can be linked to the torsion of the trajectory of the light, see~\cite{Satija2009} and~\cite{Berry1992Waveguide}. However, in our work, we considered a narrow M\"{o}bius strip where the light propagates on an undistorted and nearly planar (central) ring trajectory. Therefore, we conclude that the geometric phase arises purely from the twisting of the strip.\\
At this point, a remark on geometric phases in electronic mesoscopic transport is in order. 
The Aharonov-Bohm effect~\cite{AharonovBohm,Olariu1985} is a well-understood example. A closer relation to the present subject provide geometric phases that occur in magnetotransport through inhomogeneous magnetic fields, see e.g.~\cite{Yi1997,Frustaglia2001} and references therein. If the magnetic field is sufficiently strong (adiabatic case), the geometric phase corresponds to one half (corresponding to electronic spin) the solid angle spanned by the evolution in paramter (magnetic field) space~\cite{Berry1984}. However, for the non-adiabatic case, i.e., relatively weaker magnetic fields, the geometric phase is smaller than the original solid angle~\cite{Hentschel2004}. In all cases, the geometric phase contribution yields integer quantum numbers for vanishing magnetic field (no geometric phase), and non-integer values depending on the size of the geometric phase contribution in inhomogeneous magnetic fields. 

\begin{figure}
	\centering
	\includegraphics[width=8.5cm]{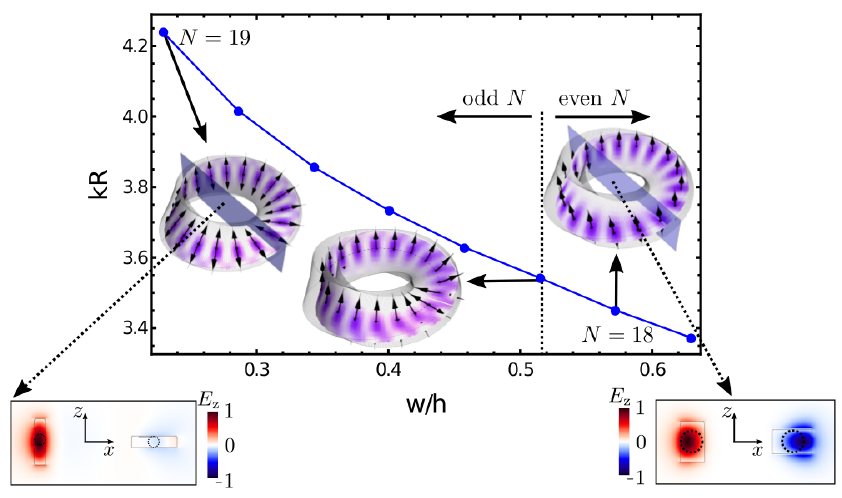}
	\caption{Resonances and polarization orientation as function of wall thickness to height ratio. Beneath a certain critical wall thickness, the polarizations remains parallel to the tube wall, and $N$ is an odd number. Above this critical value, the resonance becomes a conventional WG-type mode dominated by the torus volume inside of the strip resulting in even $N$ and constant polarization orientation. The lower insets show $x$-$z$-slices of the electric field component $E_{\text{z}}$ encoded in color scale.
	}
	\label{fig:f4}
\end{figure}

\textbf{ Manipulating the M\"{o}bius character and its topological contribution}. The question thus arises whether, similar to the electronic mesoscpic transport, 
the (so far half-integer) $m$ could be an {\it arbitrary} fractional number. When is it possible and what will happen to the resonances in this case?
To this end, we study two situations. We a) manipulate the M\"{o}bius geometry, and b) increase the thickness $w$ of the M\"{o}bius strip. 

\textit{ a) Changing the twist segment length $\Phi_{\text{TW}}$}. 
The desired manipulation can be easily achieved by reducing the extent of the segment in which the M\"{o}bius twist takes place. The standard M\"{o}bius strip is made of a strip which is twisted continuously along its length, hence the twist segment has an extent of $2\pi R$. As illustrated in Fig.~\ref{fig:f3}, we reduced the size of this segment down to the order of one wavelength. Technically, a M\"{o}bius strip is not restricted to the exact shape or structure depicted in Fig.~\ref{fig:f1}. Rather, any shape that is homeomorphic to this body is allowed. Hence, the deformed strips that we introduce here are still M\"{o}bius strips. 
If the twist segment length becomes comparable to the medium wavelength $\lambda/n$, the polarization dynamics changes as the twist segment now represents a more sudden, rather than a continuous perturbation. Accordingly, the points on the Poincar\'{e} sphere, taken again at maxima of the intensity along the M\"{o}bius strip, will remain close to the starting point $|A\rangle$. They keep the vertical polarization state because the geometry remains almost cylindrical. When the M\"{o}bius twist segment is encountered, the polarization state will change similar to the situation in the conventional M\"{o}bius strip, but now, the distance between two points on the Poincar\'{e} sphere increases as the polarization change occurs faster on a smaller portion of the trajectory. This is illustrated in Fig.~\ref{fig:f3}.  
As a consequence of Pancharatnam\textquoteright s connection, the states (points) on the Poincar\'{e} sphere have to be connected by geodesic arcs. Thus, the spanned solid angle remains $2\pi$, unless the distance of two states along the equator exceeds $\pi$. This case shows up for the first time at a segment extent of $\Phi_{\text{TW}}=32^{\circ}$, c.f. FIG.~\ref{fig:f3} b). Now, the geodesic polygon spans an angle much less than $2\pi$ and generates a geometric phase less than $\pi$, and consequently the azimuthal mode number $m$ will be some value between $9$ and $9.5$, c.f. equation~(\ref{equ:m}).

It is interesting to study the transition to non-adiabatic transport in terms of intensity. This process involves the merging of two neighboring intensity maxima and disappearance of the zero/minimum in between them, cf.~Fig.~\ref{fig:f3}b). As the geometric phase is decreased, we see a smooth transition from $N=19$ and $m=9.5$ to $N=18$ and $m=9$, cf.~Fig.~\ref{fig:f3}c). This is comparable to the smooth transition between the adiabatic and non-adiabatic situation in electronic magnetotransport. If the twist segment length is very small, it shall act like a small perturbation of a ring resonator. Indeed, we find that the resonance wave numbers do approach the one of the conventional WG mode with $m=9$ of the corresponding ring (cylinder) resonator. 

\textit{ b) Changing the wall thickness $w$}.
In a next step, we analyzed the polarization state of the resonances as a function of the wall thickness $w$ of the M\"{o}bius-strip. Interestingly, at a critical thickness $w/h$ that depends on the resonance chosen, the odd number $N$ switches to an even number, as depicted in Fig.~\ref{fig:f4}. Furthermore, the polarization then remains pointing in almost the same direction during the whole propagation cycle along the M\"{o}bius strip, and hence the waves seem to completely ignore the M\"{o}bius topology. \\
There is a simple geometric explanation for this apparent counterintuitive behavior: increasing or decreasing $w$ does not change the topology of the M\"{o}bius strip, but it changes its volume. Note that there is a torus volume inside the M\"{o}bius strip that is not affected by the twisting of the strip and changing the geometry. If the M\"{o}bius strip\textquoteright s thickness is of the order of the wavelength inside the medium, the main distribution of the electric field is located inside the invariant torus volume. As a result, a thick M\"{o}bius strip is equivalent to a (slightly perturbed) torus or cylinder.

\textbf{Far fields of the M\"{o}bius strip}.

\begin{figure}
	\centering
	\includegraphics[width=8.5cm]{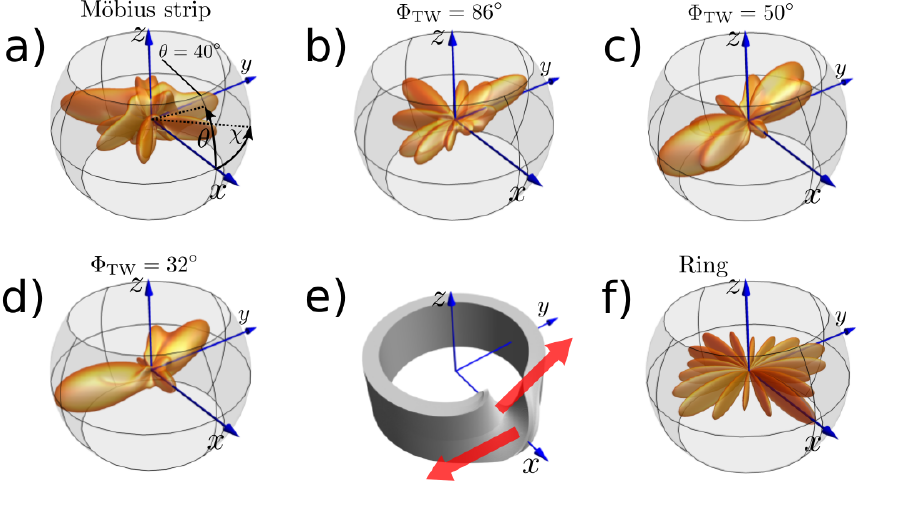}
	\caption{\textbf{a) - d)} Parametric plots of the far-field intensity of WG-like modes in the standard and modified M\"{o}bius strip cavities. \textbf{e)} Schematic illustration of the origin of bidirectional emission of a modified M\"{o}bius strip cavity. \textbf{f)} Far field of an usual ring resonator.   }
	\label{fig:f5}
\end{figure}

In the following, we investigate the far fields of the standard and the modified M\"{o}bius strips. FIG.~\ref{fig:f5} a) - d) and f) show the far-field intensity plotted as a 3D parametric plot as a function of the far-field angles $\chi$ and $\theta$ as indicated in the figure. The standard M\"{o}bius strip (c.f. FIG.~\ref{fig:f5} a) ) displays a very complex far-field emission that originates from the continues twisting of the strip wall. As the length of the twisted segment $\Phi_{\text{TW}}$ is reduced, bidirectional far-field emission emerges characterized by two main lobes along the y-axis, c.f. FIG.~\ref{fig:f5} c) and d). The origin of this bidirectional emission is the refractive output at the strongly twisted segment that represents a region of very high curvature. Most of the electromagnetic energy carried by clockwise and counter-clockwise propagating waves leaves the M\"{o}bius ring cavity at this region as schematically illustrated by two red arrows in FIG.~\ref{fig:f5} e). A remarkable  feature of all far fields presented in FIG.~\ref{fig:f5} a) - d) is that the emission is limited to the x-y-plane ($\theta=0$) because the waves propagate in an almost planar ring. For comparison, we show the far field of an usual ring resonator in FIG.~\ref{fig:f5} f) that reveals planar and isotropic emission with 18 lobes.

\textbf{Conclusions and application potential.}
 We have conducted full 3D FDTD wave calculations of thin dielectric M\"{o}bius strips in order to compute their optical modes, their polarization properties and far fields. The topological properties of the (conventional) M\"{o}bius strip result in a half-integer azimuthal mode number $m$ becasue of an extra, geometric, phase of $\pi$ provided by the Pancharatnam phase that arises from the change of the polarization state (from vertical to horizontal and back again) on the Poincar\'{e} sphere upon propagation of light through the dielectric M\"{o}bius strip. This behavior represents an example of adiabatic parallel transport of linear polarization which causes a geometric phase. Pure transverse light is governed by the same algebra as spin-$1/2$ particles, and thus the geometric phase is given by half of the solid angle spanned on the Poincar\'{e} sphere. A half-integer $m$ resonance is formed if the full upper hemisphere is spanned. 

We have presented ways to tune the value of $m$ to an arbitrary value  corresponding to a smaller solid angle enclosed on the Poincar\'{e} sphere. We have shown that this can be realized by a) decreasing the twist segment length of the M\"{o}bius strip so that the change of the polarization state occurs very fast, or by b) increasing the wall thinckness $w$. The limiting case in both situations is a coventional WG-mode resonance of the corresponding ring resonator, either because the M\"{o}bius twist is shrinked to a small perturbation as in a) or because an invariant torus formed around the central line of the M\"{o}bius strip possesses sufficient volume to guide a WG-mode as in b). In other words, we find and illustrate that, and how, topological properties can be easily manipulated via the system geometry in optical M\"{o}bius strip cavities.\\
This opens a broad venue for applications based, for example, on opto-mechanical interactions that will be useful, e.g., in sensor design. For example, when placing a dielectric 
M\"{o}bius strip on a substrate, there will be an evanescent, tapered-fiber-like coupling in the region where the M\"{o}bius strip lies flat on the substrate. This coupling will affect the cavity $Q$-factor. We find (not shown here) that this change is larger than an order of magnitude in the conventional M\"{o}bius geometry, and much smaller in M\"{o}bius geometries with very small twist segment lengths, such that a change in the cavity $Q$-factor could be directly related to the system geometry and mechanical properties. This will be the subject of further studies. Our findings might also be useful in seperating the geometric and spin-orbit-coupling contributions in the polarization evolution in more complex systems such as in~\cite{Ma2016}.

\begin{acknowledgments}
	This work was partly supported by Emmy-Noether programme of the German Research Foundation (DFG).
\end{acknowledgments}


%

\end{document}